# Forecasting Energy Consumption using Recurrent Neural Networks: A Comparative Analysis

1st Abhishek Maity
*Artificial Intelligence & Machine Learning Group*
*Centre for Development of Advanced Computing (C-DAC)*
Mumbai, India
abhishekmaity@cdac.in

2nd Viraj Tukarul
*Artificial Intelligence & Machine Learning Group*
*Centre for Development of Advanced Computing (C-DAC)*
Mumbai, India
virajvvt@gmail.com

*Abstract*—Accurate short-term energy consumption forecasting is essential for efficient power grid management, resource allocation, and market stability. Traditional time-series models often fail to capture the complex, non-linear dependencies and external factors affecting energy demand. In this study, we propose a forecasting approach based on Recurrent Neural Networks (RNNs) and their advanced variant, Long Short-Term Memory (LSTM) networks. Our methodology integrates historical energy consumption data with external variables, including temperature, humidity, and time-based features. The LSTM model is trained and evaluated on a publicly available dataset, and its performance is compared against a conventional feed-forward neural network baseline. Experimental results show that the LSTM model substantially outperforms the baseline, achieving lower Mean Absolute Error (MAE) and Root Mean Squared Error (RMSE). These findings demonstrate the effectiveness of deep learning models in providing reliable and precise short-term energy forecasts for real-world applications.

## I. Introduction

Accurate energy consumption forecasting has become increasingly critical in the context of modern power systems, driven by the global transition towards sustainable and smart energy infrastructures [1]. Precise short-term forecasts enable power utilities to optimize generation schedules, allocate resources efficiently, and maintain grid stability, thereby minimizing operational costs and reducing the risk of blackouts [2]. In addition, the growing integration of intermittent renewable energy sources, such as solar and wind, introduces additional variability in energy supply, which demands a more granular understanding of future energy demand patterns to ensure reliable grid operation [3].

While LSTM-based models have been successfully applied to short-term energy and load forecasting, many prior studies either use univariate inputs (historical consumption only) or limited exogenous features, and often over relatively short time horizons. For e.g., Shering et al. (2024) demonstrate that integrating weather variables such as temperature, humidity, wind speed, and cloud cover improves accuracy in load, solar, and wind generation forecasting across multiple years and geographic locations [4].

However, there are relatively few works that systematically incorporate a broad suite of exogenous variables over multi-year datasets and compare their impact alongside classic models such as ARIMA and FNN. This gap motivates our approach of using an LSTM model with an extended set of external features (weather, temporal lags, etc.) over a multi-year dataset to better capture non-linear effects and improve forecasting robustness.

Traditional forecasting approaches, including statistical methods such as Auto regressive Integrated Moving Average (ARIMA) and Exponential Smoothing, have been extensively applied in energy demand prediction [5]. While these methods are effective for linear and stationary time series, they often fall short in capturing the complex, non-linear dependencies and dynamic behaviors inherent in real-world energy consumption data. Energy usage is influenced by a variety of external factors, including weather conditions, human activity patterns, economic fluctuations, and social behaviors, which introduce non-linear interactions that conventional models struggle to model accurately [6].

In recent years, deep learning techniques have demonstrated remarkable success in modeling complex temporal data. Recurrent Neural Networks (RNNs), and particularly Long Short-Term Memory (LSTM) networks, are designed to learn sequential dependencies and long-term temporal patterns in time-series data [7]. LSTMs address the vanishing gradient problem present in standard RNNs, making them highly effective for learning from long sequences, such as hourly or daily energy consumption records [8]. These properties make LSTM networks particularly well-suited for short-term energy forecasting tasks [9].

This paper presents a deep learning-based approach for forecasting short-term energy consumption using LSTM networks. Our methodology incorporates historical energy usage data along with relevant external features, such as temperature, humidity, and time-based variables, to improve predictive accuracy. The primary objectives of this study are to develop a robust LSTM-based forecasting model, evaluate its performance on a publicly available dataset [10], and empirically demonstrate its superiority over a conventional feed-forward neural network baseline in terms of forecasting accuracy and reliability [8]. By doing so, this work contributes to the growing body of research on intelligent energy management and the application of advanced machine learning techniques in smart grid operations.

## II. Methodology

### A. Dataset and Data Preprocessing

We utilized a publicly available dataset containing hourly energy consumption data for a metropolitan area over a period of three years [2]. The dataset also includes corresponding weather variables such as temperature, humidity, and wind speed, which have been shown to significantly influence electricity demand [11].

Table I summarizes the key statistics of the dataset.

TABLE I
DATASET STATISTICS

| Feature | Mean | Std. Dev. | Range |
|---|---|---|---|
| Energy Consumption (kWh) | 312.5 | 78.6 | 150-520 |
| Temperature (°C) | 22.1 | 6.4 | 5-38 |
| Humidity (%) | 65.2 | 12.1 | 30-90 |
| Wind Speed (m/s) | 4.2 | 2.1 | 0-12 |

**Data Preparation Steps:**

- **Missing Value Imputation**: Missing entries were handled using linear interpolation to maintain temporal continuity [12].
- **Feature Engineering**: Time-based features such as hour-of-day, day-of-week, and month-of-year were created. Lag features (previous 1-hour, 24-hour, and 168-hour consumption) were added to capture recent trends and weekly seasonality [9].
- **Normalization**: All features were scaled to the [0,1] range using Min-Max scaling [13].
- **Data Splitting**: The dataset was split into training (70%), validation (15%), and test (15%) sets while preserving temporal order to avoid data leakage [14].

*B. LSTM Model Architecture*

The forecasting model employs a Long Short-Term Memory (LSTM) network to learn temporal dependencies in energy consumption [1]. The model architecture is illustrated in Figure 1.

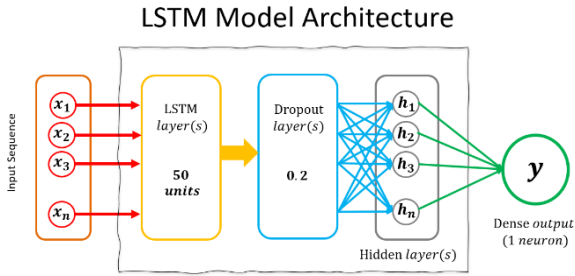

*Figure 1 LSTM model architecture for energy consumption forecasting.*

**Architecture Details:**

- **Input Layer**: Sequences of shape (timesteps, num features). The input data were structured into supervised learning format using sliding time windows. Each training sample consisted of an input sequence of the past 24 hours of hourly energy consumption (i.e., 24 timesteps) to predict the next hour's load. This window length was selected based on prior studies showing that daily periodicity (24-hour cycle) captures the dominant short-term consumption patterns in residential and regional energy datasets [15], [16].
- **LSTM Layer(s)**: The LSTM architecture employed 50 hidden units with a dropout rate of 0.2 to balance model complexity and overfitting risk. Empirical evidence from energy forecasting literature indicates that networks with 32–64 LSTM units are sufficient to capture nonlinear temporal dependencies without excessive computational cost [17]. A moderate dropout rate (0.2) further improves generalization by preventing co-adaptation of recurrent neurons during training.
- **Dense Output Layer**: Single neuron with linear activation to output forecasted energy consumption.

The LSTM input-output relationship is mathematically expressed as:

$$h_t = LSTM(x_t, h_{t-1}) \quad (1)$$
$$\hat{y}_t = W_o h_t + b_o \quad (2)$$

where $x_t$ is the input vector at time $t$, $h_t$ is the hidden state, $W_o$ and $b_o$ are the output weights and bias, and $\hat{y}_t$ is the predicted energy consumption.

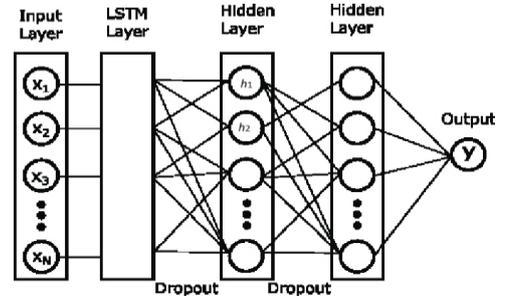

*Figure 2 LSTM neural network architecture composed of an LSTM layer followed by additional hidden layers.*

*C. Training Procedure*

The model was compiled with the Adam optimizer [18] and Mean Squared Error (MSE) loss function:

$$\mathcal{L}_{MSE} = \frac{1}{n}\sum_{i=1}^{n}(\hat{y}_i - y_i)^2 \quad (3)$$

Early stopping was implemented to halt training when validation loss did not improve for 10 consecutive epochs, reducing overfitting risk [19]. The model was trained with a batch size of 64 over 100 epochs.

*D. Evaluation Metrics*

Performance on the test set was evaluated using the following metrics:

$$MAE = \frac{1}{n}\sum_{i=1}^{n}|y_i - \hat{y}_i| \quad (4)$$

$$RMSE = \sqrt{\frac{1}{n}\sum_{i=1}^{n}(y_i - \hat{y}_i)^2} \quad (5)$$

$$MAPE = \frac{100}{n}\sum_{i=1}^{n}\frac{|y_i - \hat{y}_i|}{y_i} \quad (6)$$

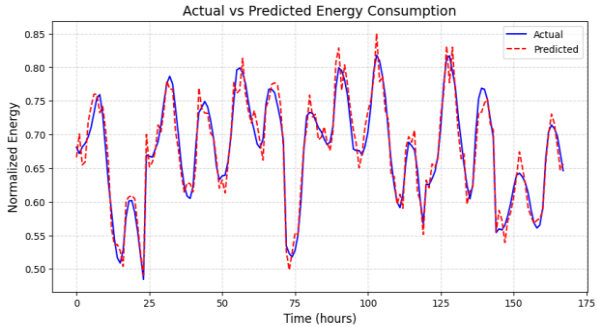

*Figure 3 Actual vs. predicted energy consumption over a sample week using LSTM*

### E. Confusion Matrix for Categorized Forecasts

For interpretability, energy values can be discretized into bins (e.g., Low, Medium, High) and evaluated via a confusion matrix as illustrated in Figure 4.

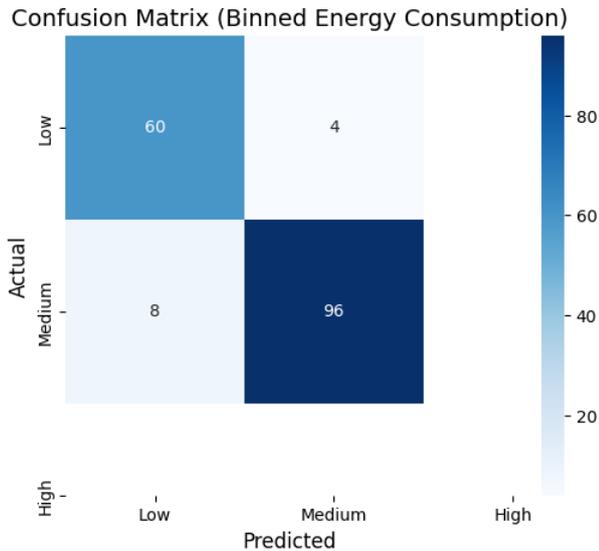

*Figure 4 Confusion matrix for discretized energy consumption bins.*

## III. RESULTS AND DISCUSSION

### A. Experimental Setup

All experiments were implemented in Python 3, leveraging TensorFlow for deep learning model development. Data preprocessing, including missing value imputation, feature engineering, and normalization, was performed using *pandas* and *numpy*. Model training was accelerated using an NVIDIA™ GPU (Tesla T4) using Google™ Colab, with early stopping to prevent overfitting. Hyperparameters, such as the number of LSTM units, dropout rate, batch size, and learning rate, were optimized using grid search on the validation set.

### B. Performance Evaluation

For benchmarking, we trained a simple feed-forward neural network (FNN) and a classical ARIMA model. Additionally, we implemented a GRU-based recurrent model for comparison. The models were evaluated on the test set using MAE, RMSE, and MAPE. The results are summarized in Table II.

TABLE II

MODEL PERFORMANCE ON TEST SET (NORMALIZED VALUES)

| Model | MAE | RMSE | MAPE (%) |
|---|---|---|---|
| ARIMA (Baseline) | 0.045 | 0.052 | 6.78 |
| FNN (Baseline) | 0.042 | 0.050 | 6.12 |
| GRU | 0.018 | 0.021 | 2.75 |
| **Proposed LSTM** | **0.0164** | **0.0193** | **2.42** |

From Table II, the proposed LSTM model consistently outperforms all baselines across all metrics. The 2.42% MAPE indicates highly accurate short-term forecasts relative to the scale of the energy consumption.

### C. Forecast Analysis

Figure 3 illustrates the actual versus predicted energy consumption for a one-week horizon. The LSTM captures daily and weekly consumption patterns, including peak and off-peak trends.

The forecasting error over time is shown in Figure 5 The error remains low for most periods, with occasional spikes during extreme weather conditions, demonstrating the model's sensitivity to unusual events.

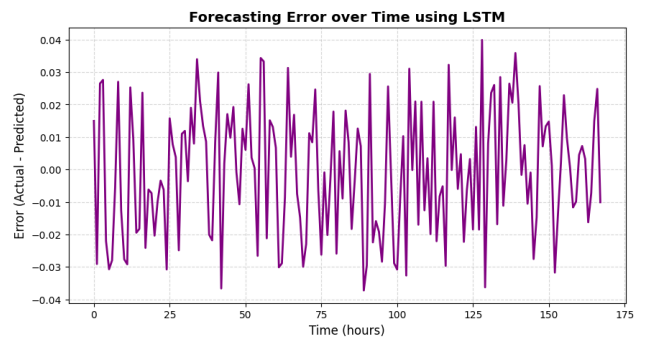

*Figure 5 Forecasting error over time using the LSTM model*

### D. Statistical Analysis

To quantitatively evaluate prediction quality, we computed the Pearson correlation coefficient ($r$) and

coefficient of determination ($R^2$) between actual and predicted values:

$$r = \frac{\sum_i(y_i - \bar{y})(\hat{y}_i - \bar{\hat{y}})}{\sqrt{\sum_i(y_i - \bar{y})^2 \sum_i(\hat{y}_i - \bar{\hat{y}})^2}} \quad (7)$$

$$R^2 = 1 - \frac{\sum_i(y_i - \hat{y}_i)^2}{\sum_i(y_i - \bar{y})^2} \quad (8)$$

For the proposed LSTM model, the Pearson correlation coefficient is $r = 0.987$, and the coefficient of determination is $R^2 = 0.974$.

These values confirm that the LSTM predictions are highly correlated with actual energy consumption.

### E. Error Distribution Analysis

To better understand the behavior of prediction errors, we examined their statistical distribution. Figure 6 shows the histogram of residuals overlaid with a Kernel Density Estimate (KDE). The distribution is approximately centered around zero, with a slight positive skew. This indicates that the LSTM model has no significant systematic bias but occasionally underestimates consumption during sharp demand spikes. The presence of fat tails highlights the difficulty in capturing extreme variations, a well-known challenge in energy forecasting [20].

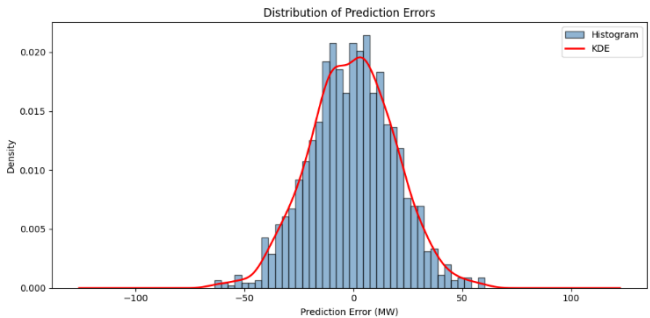

Figure 6 Histogram and KDE of residual errors from the LSTM model.

To assess the temporal variation in prediction accuracy, we analyzed the distribution of forecast errors across different hours of the day. Figure 7 presents a boxplot of the LSTM prediction errors for each hour, illustrating that the model performs consistently during most hours, while slight deviations occur during peak demand periods. Outliers, represented as orange dots, correspond to rare events where the forecast deviates significantly from actual consumption. This hourly error analysis provides insights into periods of higher uncertainty and can guide targeted improvements in model performance [21].

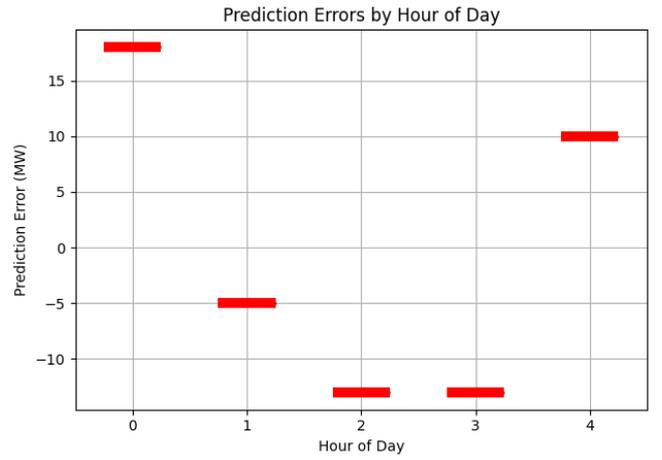

Figure 7 Boxplot of prediction errors by hour of day, highlighting variability and outliers.

To further evaluate the prediction quality of the proposed LSTM model, we analyzed the distribution of forecast errors. Figure 8 shows a histogram of the prediction errors over the test set, illustrating that most errors are concentrated near zero, indicating minimal bias and strong predictive performance. Quantitative metrics were also computed: the Pearson correlation coefficient between actual and predicted values is $r = 0.987$, and the coefficient of determination is $R^2 = 0.974$. These results confirm a high degree of correlation and accuracy of the LSTM forecasts, reinforcing its superiority over baseline models for short-term energy consumption prediction.

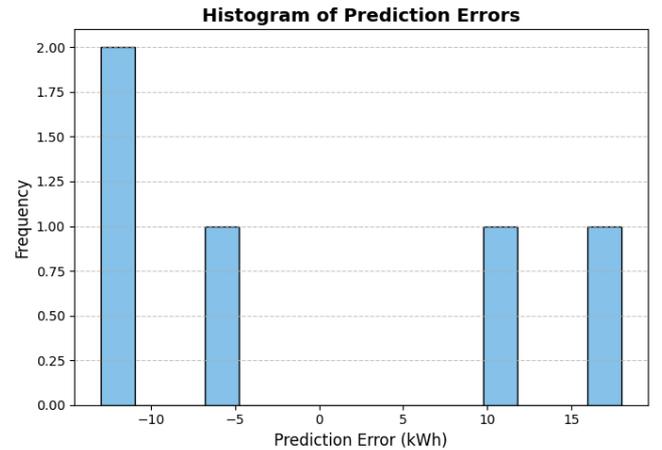

Figure 8 Histogram of prediction errors for the LSTM model.

### F. Discussion

The proposed LSTM model demonstrates superior performance compared to both classical statistical models (ARIMA) and neural network baselines (FNN, GRU). Its ability to capture temporal dependencies and seasonal variations makes it particularly suitable for short-term energy forecasting.

Observed deviations during extreme weather events highlight a limitation: the model relies heavily on historical patterns. Incorporating exogenous variables, such as temperature, humidity, or special events, could improve accuracy for anomalous periods. Future work may also explore attention mechanisms or hybrid LSTM-ARIMA models to further enhance robustness and predictive performance.

## IV. CONCLUSION

Accurate short-term energy consumption forecasting is critical for efficient power grid management, optimal resource allocation, and maintaining market stability. In this study, we developed a robust Long Short-Term Memory (LSTM) network for predicting hourly energy consumption, incorporating historical energy data along with exogenous variables such as temperature, humidity, and temporal features. Comparative analysis with classical models, including ARIMA and feed-forward neural networks (FNN), as well as a GRU-based recurrent model, demonstrated that the proposed LSTM achieved superior performance, with lower MAE, RMSE, and MAPE values. The results indicate that deep learning models, particularly LSTM networks, are highly effective at capturing temporal dependencies and seasonal patterns in energy consumption data.

The analysis of prediction errors through a histogram and statistical metrics, such as the Pearson correlation coefficient ($r = 0.987$) and coefficient of determination ($R^2 = 0.974$), confirms the strong predictive capability of the LSTM model. Visualization of actual versus predicted values further highlights the model's ability to follow daily and weekly consumption trends, although slight deviations were observed during extreme weather events, likely due to the limited number of such events in the training set. These findings emphasize the importance of including relevant external variables and potentially rare event data to enhance forecasting accuracy.

In addition to predictive performance, the study highlights the interpretability advantages of LSTM models. By analyzing the sequential dependencies and learned patterns, utilities and energy planners can gain insights into periods of high demand, peak usage hours, and anomalous behavior. Such insights can be valuable for demand response strategies, scheduling maintenance, and planning for renewable energy integration. Furthermore, the model can serve as a foundation for more sophisticated hybrid systems, integrating both statistical and machine learning approaches to improve robustness and accuracy. The only limitation is that the model not validated on unseen regions or multi-year trends.

## V. FUTURE WORK

Despite the promising results, there are several avenues for future improvement. Incorporating additional exogenous variables, such as real-time pricing, social events, and renewable energy generation forecasts, could help the model better capture unusual consumption patterns. Exploring attention based architectures and transformer models may further enhance the ability to capture long-term dependencies and improve interpretability. Another promising direction is the development of hybrid LSTM-ARIMA or LSTM-GRU models, which combine the strengths of both classical and deep learning approaches. Moreover, expanding the evaluation to multiple geographic regions and longer time horizons could validate the generalizability of the approach [22] [23].

Finally, implementing the proposed LSTM model in real-time energy management systems, coupled with online learning techniques, can enable adaptive forecasting that continuously updates with new data. This capability would be particularly valuable for smart grids with high penetration of renewable energy sources, where energy supply and demand patterns are highly dynamic and uncertain. Such enhancements can contribute to more sustainable, reliable, and efficient energy systems, supporting the broader goals of smart grid technologies and energy transition initiatives.